\documentclass[reprint,superscriptaddress,showpacs,preprintnumbers,amsmath,amssymb,aps,prd]{revtex4-1}

\usepackage{graphicx}
\usepackage{dcolumn}
\usepackage{bm}
\usepackage{color}

\usepackage{cancel}
\usepackage{multirow}
\newcommand{\met}{\cancel{E}_T}

\begin{document}

\preprint{DESY 16-202}

\title{Search for Heavy Sterile Neutrinos in Trileptons at the LHC}

\author{Claudio O. Dib}
\email{claudio.dib@usm.cl}
\affiliation{ CCTVal and Department of Physics, Universidad T\'ecnica Federico Santa Mar{\'{\i}}a, Valpara{\'{\i}}so, Chile }
\author{C. S. Kim}%
\email{cskim@yonsei.ac.kr}
\affiliation{ Department of Physics and IPAP, Yonsei University, Seoul 120-749, Korea }%

\author{Kechen Wang}
\email{kechen.wang@desy.de (corresponding author)}
\affiliation{ DESY, Notkestra횩e 85, D-22607 Hamburg, Germany }%
\affiliation{ Center for Future High Energy Physics, Institute of High Energy Physics, Chinese Academy of Sciences, Beijing, 100049, China }%


\begin{abstract}
\noindent We present a search strategy for both Dirac and Majorana sterile neutrinos from the purely leptonic decays of $W^\pm \to e^\pm  e^\pm  \mu^\mp \nu$ and $\mu^\pm  \mu^\pm  e^\mp \nu$ at the 14 TeV LHC. The discovery and exclusion limits for sterile neutrinos are shown using both the Cut-and-Count (CC) and Multi-Variate Analysis (MVA) methods.
We also discriminate between Dirac and Majorana sterile neutrinos by exploiting a set of kinematic observables which differ between the Dirac and Majorana cases.
We find that the MVA method, compared to the more common CC method, can greatly enhance the discovery and discrimination limits. Two benchmark points with sterile neutrino mass $m_N = 20$ GeV and 50 GeV are tested.
For an integrated luminosity of 3000 ${\rm fb}^{-1}$, sterile neutrinos can be found with $5 \sigma$ significance if heavy-to-light neutrino mixings $|U_{Ne}|^2 \sim |U_{N\mu}|^2\sim 10^{-6}$, while Majorana vs. Dirac discrimination can be reached if at least one of the mixings is of order $10^{-5}$.
\end{abstract}


\maketitle

\noindent{\bf Introduction} The evidence of small but non zero neutrino masses \cite{Oscillation_experiment} is currently an outstanding path
beyond the Standard Model of particle physics. Most explanations are based on the existence of extra heavy particles. In particular, seesaw models involve extra heavy neutrinos that are sterile under electroweak interactions, but which mix with the Standard leptons \cite{seesaw_models}. Moreover, in most scenarios they are Majorana fermions \cite{Majorana_theory}.
The existence of heavy neutrinos and the discrimination between Dirac and Majorana is thus a crucial piece of information that experiments must reveal.
The Majorana nature of neutrinos is searched in neutrinoless double beta decays~\cite{Engel:2016xgb},
but so far no experimental evidence has been found~\cite{Majorana_experiments}.
The Large Hadron Collider (LHC) and future colliders also offer the opportunity to search for heavy neutrinos \cite{Deppisch:2015qwa,Antusch:2016ejd}.
At such colliders, same-sign dilepton plus dijet events, $\ell^\pm\ell^\pm j j$, can be produced if there are heavy Majorana neutrinos (henceforth called $N$)  in the intermediate state with masses above $M_W$ \cite{dileptonLHC}. Instead, for masses below $M_W$, the jets are lost in the background and thus trilepton events $\ell^\pm\ell^\pm\ell^{\prime\mp}\nu$ provide clearer signals for a heavy $N$~\cite{Izaguirre:2015pga},
where $\ell$ and $\ell^{\prime}$ denote leptons with different flavors.
The choice of having no Opposite-Sign Same-Flavor (no-OSSF) lepton pairs helps eliminate a serious SM background  $\gamma^*/Z \to \ell^+\ell^-$ \cite{p2mu}.
Now, if $N$  is Majorana, the trilepton will contain a Lepton Number Conserving (LNC) channel $W^+ \to e^+ e^+ \mu^- \nu_e$  as well as a Lepton Number Violating (LNV) channel $W^+ \to e^+ e^+ \mu^- \bar\nu_\mu$, while if it is of Dirac type, only the LNC channel will appear.
An in-between case of neutrino called \emph{pseudo-Dirac} occurs if $N$ corresponds to pairs of almost degenerate Majorana neutrinos so that the LNV mode becomes relatively suppressed by two interfering amplitudes \cite{Anamiati:2016uxp}. Here we will not consider such a case.
Since the final neutrino escapes the detection, the observed final state is just $e^\pm e^\pm \mu^\mp$ or $\mu^\pm \mu^\pm e^\mp$ { plus missing energy. Hence it is not a simple task} to distinguish a  Majorana vs. a Dirac $N$.
In our previous work \cite{Dib:2015oka}, we studied these trilepton events to discover heavy neutrinos and discriminate between Dirac and Majorana using differences in their energy distributions.
In our consecutive work \cite{Dib:2016wge}, we presented a simpler method
for this discrimination
by comparing the full rates of $e^\pm e^\pm \mu^\mp$ and $\mu^\pm \mu^\pm e ^\mp$.
 However,
this discrimination based on full rates only works if  the mixing parameters $U_{Ne}$ and $U_{N\mu}$ are considerably different from each other (See Table 1).
\\

\noindent{\bf Discovery Limit:} In this letter, we present a strategy to discover heavy sterile neutrinos $N$ with $m_N < M_W$, and discriminate between their Dirac vs. Majorana character, using trilepton events at the 14 TeV LHC, applying both a Cut-and-Count (CC) and a Multi-Variate Analysis (MVA) methods. Our strategy is most complete in the sense that uses all details of each event, including spectra and angular distributions.

\begin{figure}
\includegraphics[scale=0.7]{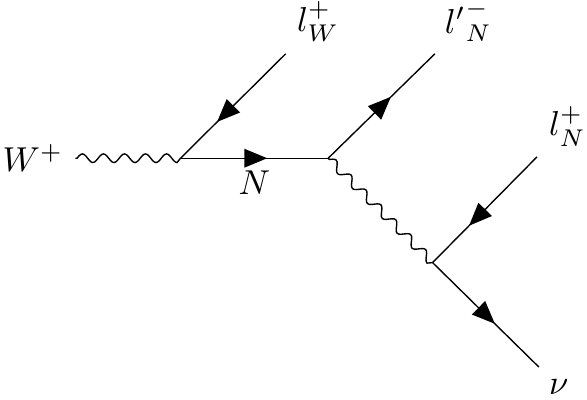}
\includegraphics[scale=0.7]{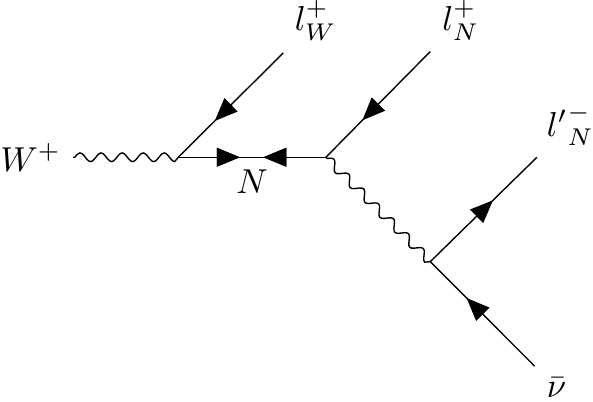}
\caption{ The LNC process $W^+\to l_W^+ {l^\prime}_N^- l_N^+ \nu$, mediated by a heavy sterile neutrino of Majorana or Dirac type (left); and the LNV process $W^+ \to l_W^+ l_N^+ {l^\prime}_N^- \bar\nu$,  mediated by a heavy sterile neutrino of Majorana type (right);  }
\label{fig:Wdecay}
\end{figure}

We consider the process $W^{\pm} \to l_{W}^{\pm} l_{N}^{\pm} {l^\prime}_{N}^{\mp} \nu$
(Fig.~\ref{fig:Wdecay}), where $l$ and $l^\prime$ are different leptons, either $e$ or $\mu$ (i.e. $e^\pm  e^\pm  \mu^\mp \nu$ and $\mu^\pm  \mu^\pm  e^\mp \nu$),
 and $\nu$ is a SM neutrino or antineutrino.
For convenience, we introduce two parameters: a \emph{normalization} factor $s$ and a \emph{disparity} factor $r$:
\begin{equation}
s \equiv 2\times10^6\, \frac{|U_{Ne} U_{N\mu} |^2}{|U_{Ne}|^2+|U_{N\mu}|^2} ,\,\,\,\,\, r \equiv \frac{|U_{Ne}|^2}{|U_{N\mu}|^2} .
\label{sr}
\end{equation}
Conversely, the heavy-to-light mixing elements $|U_{Ne}|^2$ and $|U_{N\mu}|^2$ can be expressed in terms of $r$ and $s$ as:
\begin{equation}
|U_{Ne}|^2 = \frac{s\, (1 + r)}{2 \times 10^6},\quad  |U_{N\mu}|^2 = \frac{s\, (1 + \frac{1}{r} )}{2 \times 10^6}.
\label{eqn:mixings}
\end{equation}

For our study we choose two benchmark points: $m_N = $ 20 and 50 GeV, with $r = s = 1$
(i.e., $|U_{Ne}|^2=|U_{N\mu}|^2=10^{-6}$).
The production rates of the different trilepton modes are proportional to the scale factors shown in Table~\ref{tab:Nsig}.

\begin{table}[h]
\centering
\begin{tabular}{ccc}
\hline
\hline
  & Dirac & Majorana   \\
\hline
$e^{\pm} e^{\pm} \mu^{\mp}$ & $s$ & $s\,(1+r)$ \\
$\mu^{\pm} \mu^{\pm} e^{\mp}$ & $s$ & $s\,(1+ 1/r )$ \\
\hline
\hline
\end{tabular}
\caption{Scale factors for the production rates of the trilepton final states. See Eq.~(\ref{sr}) for the definitions
of $s$ and $r$.
}
\label{tab:Nsig}
\end{table}

Let us first describe our strategy to discover or set exclusion limits for Dirac and Majorana sterile neutrinos using trileptons at the LHC.
We first select trilepton events $l^{\pm} l^{\pm} {l^\prime}^{\mp}$ with no-OSSF lepton pairs.
Then we apply basic cuts for leptons and jets: $p_{T,l} \geq$ 10 GeV and $|\eta_l| \leq$ 2.5; $p_{T,j} \geq$ 20 GeV and $|\eta_j| \leq$ 5.0, and veto the b-jets in order to suppress the $t\bar{t}$ background.
Now, in order to select within the pair $l^{\pm} l^{\pm}$ the lepton that comes from the $N$ decay, we construct the $\chi^2$ function
\begin{equation}
\chi^2= ( {\cal M}_W - m_W)^2/\sigma_W^2 + ( {\cal M}_N - m_N)^2/\sigma_N ^2,
\label{eqn:Chi2}
\end{equation}
where $m_W=80.5$ GeV  and $m_N$ is the assumed mass for $N$ (20 or 50 GeV in our benchmarks), while ${\cal M}_W$ and ${\cal M}_N$ are the reconstructed invariant masses of  $l^{\pm} l^{\pm} {l^\prime}^{\mp} \nu$ and $l^{\pm} {l^\prime}^{\mp} \nu$, respectively; $\sigma_W$ and $\sigma_N$ are the widths of the reconstructed mass distributions, which we take to be 5\% of their respective  $m_W$ and $m_N$, for simplicity. When calculating the reconstructed mass $M_W$ and $M_N$, the final neutrino transverse momentum
${\textbf{p}}_{T,\nu}$
is assumed to be the missing transverse momentum, while the neutrino longitudinal momentum $p_{z,\nu}$ and the correct lepton $l^{\pm}$ from the $N$ decay are determined by minimizing the $\chi^2$ of Eq.~(\ref{eqn:Chi2}).

A better identification of the correct lepton can be achieved if the production and decay vertices of $N$ are spatially displaced in the detector \cite{Helo:2013esa, Dib:2014iga}. However, this would be perceptible only if $m_N \lesssim 15$ GeV at the LHC.
For $m_N \sim 15$ GeV, by exploiting the displaced lepton jet search and requiring the vertex displacement between 1 mm and 1.2 m, the Ref.~\cite{Izaguirre:2015pga} derived a limits of $|U_{N\mu}|^2 < 10^{-5}$ at 8 TeV LHC with 20 ${\rm fb}^{-1}$, and $|U_{N\mu}|^2 < 10^{-7}$ at 13 TeV LHC with 300 ${\rm fb}^{-1}$ at 2-$\sigma$ level.
A future $e^-e^+$ collider with better detector resolution of the vertex displacement will allow to probe heavier sterile neutrinos. By requiring the vertex displacement between 10 $\rm {\mu m}$ and 249 cm at the FCC-ee, the Ref.~\cite{Antusch:2016vyf} yields the sensitivity of $|U_{Nl}|^2 \sim 10^{-11}$ for the Z-pole running mode with 110 ${\rm fb}^{-1}$, and the sensitivity of $|U_{Ne}|^2 \sim 10^{-8}$ for a 240 GeV running with 5 ${\rm ab}^{-1}$ at 2-sigma level.
Due to the much more challenging experimental environment, the sensitivity at the FCC-hh might not be as good as that from the FCC-ee.
For this study, the displaced vertex observable is not considered.

A MVA is then performed to exploit the useful observables and maximally reduce the SM background. We use the \emph{Boosted Decision Trees} (BDT) method in the TMVA package~\cite{TMVA2007} and input the following kinematical observables for training and test processes:
($i$) the missing energy $\met$;
($ii$) the scalar sum of $p_T$ of all jets $H_T$;
($iii$) the transverse mass of the missing energy plus lepton(s)
$M_T(\met, l_W l_N {l^\prime}_N)$,
$M_T(\met, l_N {l^\prime}_N)$,
$M_T(\met, l_W {l^\prime}_N)$,
$M_T(\met, l_W)$,
$M_T(\met, l_N)$,
$M_T(\met, {l^\prime}_N)$;
($iv$) the azimuthal angle difference $\Delta\phi$ between the missing transverse momentum and lepton(s)
$\Delta\phi(\met, l_N {l^\prime}_N)$,
$\Delta\phi(\met, l_W {l^\prime}_N)$,
$\Delta\phi(\met, l_W)$,
$\Delta\phi(\met, l_N)$,
$\Delta\phi(\met, {l^\prime}_N)$;
($v$) the invariant mass of the system of leptons
$M(l_W l_N {l^\prime}_N)$,
$M(l_W l_N)$,
$M(l_W {l^\prime}_N)$,
$M(l_N {l^\prime}_N)$;
and ($vi$) the azimuthal angle difference $\Delta\phi$ between two leptons
$\Delta\phi(l_W, {l^\prime}_N)$,
$\Delta\phi(l_N, {l^\prime}_N)$.
For a Dirac (Majorana) $N$, the simulation data of the LNC (LNC + LNV) processes are inputs as the signal sample, while the total SM background data ($\gamma^*/Z$, WZ, and $t\bar{t}$ inclusively) are inputs as the background sample for the TMVA training and test processes.
The details of our data simulation procedures are described in \cite{Dib:2016wge}.

\begin{figure}[h]
\includegraphics[scale=0.09]{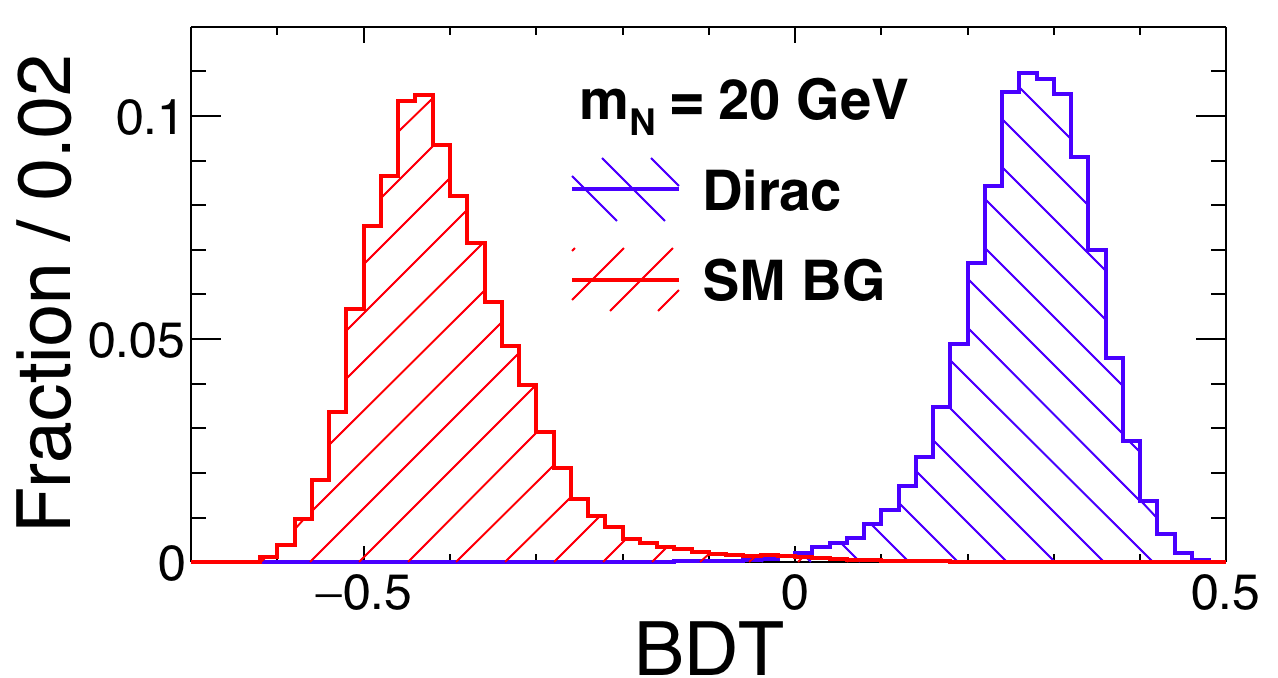}
\includegraphics[scale=0.09]{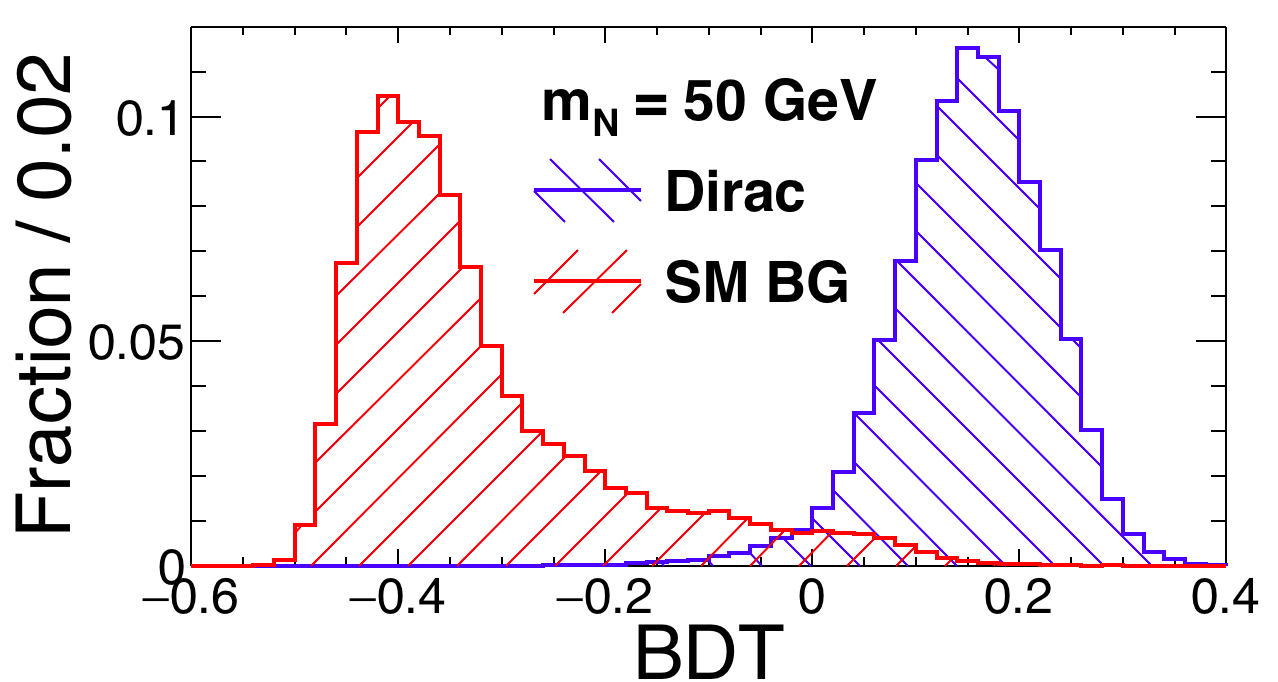}
\caption{ Distributions of BDT response for Dirac signal (blue) with $m_N$ = 20 (left) and 50 (right) GeV, and total SM backgrounds (red) including $\gamma^*/Z$+jets, WZ+jets and $t\bar{t}$.}
\label{fig:BDTdrcVSbg}
\end{figure}

Fig.~\ref{fig:BDTdrcVSbg} shows the BDT response distributions for a Dirac $N$ signal and total SM background, for our two benchmarks. The signal vs. background separation is better for $m_N = $ 20 GeV than for $m_N = $ 50 GeV, as the two curves have less overlap in Fig.~\ref{fig:BDTdrcVSbg} (left).

In Table~\ref{tab:sigVSbg_mN20}, we show the number of events for both Dirac and Majorana signals with $m_N = 20$ GeV and the SM backgrounds at the 14 TeV LHC. The first two rows show the number of events after basic cuts and b-jets vetoes. The number of events using the CC method from Ref.~\citep{Dib:2016wge} are shown in the third row. The numbers of events for Dirac (Majorana) sterile neutrinos using the BDT method are shown in the fourth (fifth) row.
For a
Dirac (Majorana) $N$,
we get a statistical significance
\begin{equation}
\mathit{SS} = N_s/\sqrt{N_s+N_b}
\end{equation}
near 2.6 (5.8) for the CC method and near 6.6 (10.7) for the BDT method, where $N_s$ and $N_b$ are the number of signal events (either Dirac or Majorana) and SM background events, respectively.
Similarly, Table~\ref{tab:sigVSbg_mN50} shows the numbers for $m_N = 50$ GeV. From Fig.~\ref{fig:BDTdrcVSbg}, lower significances are expected for $m_N = 50$ GeV. Indeed, Table~\ref{tab:sigVSbg_mN50} shows SS near 2.3 (4.8) for the CC method and near 5.1 (9.0) for the BDT method.

\begin{table}[h]
\centering
\begin{tabular}{ccccccc}
\hline
\hline
Cuts            & Dirac & Majorana & $\gamma^*/Z$ & WZ & $t\bar{t}$ & $\mathit{SS}$   \\
Basic cuts         & 54.0 & 133.2 & 4220   & 2658   & 68588   \\
N(b-jets)=0        & 53.1 & 131.1 & 4063.0 & 2497.1 & 31953.5 \\
\hline
CC                 & 44.2 & 110.9 &  209.8 &   25.3 &    16.9 &  2.6 (5.8)  \\
${\rm BDT}>0.183$ & 46.7 &    -  &    1.9 &    1.3 &     0.0 &  6.6 \\
${\rm BDT}>0.171$ &   -  & 120.7 &    5.1 &    1.7 &     0.8 & 10.7 \\
\hline
\hline
\end{tabular}
\caption{Cut flow for signal and background processes with $m_N^{} = 20~\mathrm{GeV}$. Numbers of events correspond to an integrated luminosity of $3000~\mathrm{fb}^{-1}$ at the $14~\mathrm{TeV}$ LHC.}
\label{tab:sigVSbg_mN20}
\end{table}

\begin{table}[h]
\centering
\begin{tabular}{ccccccc}
\hline
\hline
Cuts            & Dirac & Majorana & $\gamma^*/Z$ & WZ & $t\bar{t}$ & $\mathit{SS}$ \\
Basic cuts         & 108.4 & 228.8 & 4220   & 2658   & 68588 &   \\
N(b-jets)=0        & 106.7 & 225.2 & 4063.0 & 2497.1 & 31953.5 & \\
\hline
CC                 &  91.9 & 193.9 & 1283.1 &  120.7 &    48.9 & 2.3 (4.8) \\
${\rm BDT}>0.138$ &  64.4 &    -  &   25.7 &   47.5 &    21.1 & 5.1 \\
${\rm BDT}>0.138$ &    -  & 143.2 &   31.0 &   52.8 &    27.0 & 9.0 \\
\hline
\hline
\end{tabular}
\caption{Cut flow for signal and background processes with $m_N^{} = 50~\mathrm{GeV}$. Numbers of events correspond to an integrated luminosity of $3000~\mathrm{fb}^{-1}$ at the $14~\mathrm{TeV}$ LHC.}
\label{tab:sigVSbg_mN50}
\end{table}

\begin{figure}[h]
\includegraphics[scale=0.09]{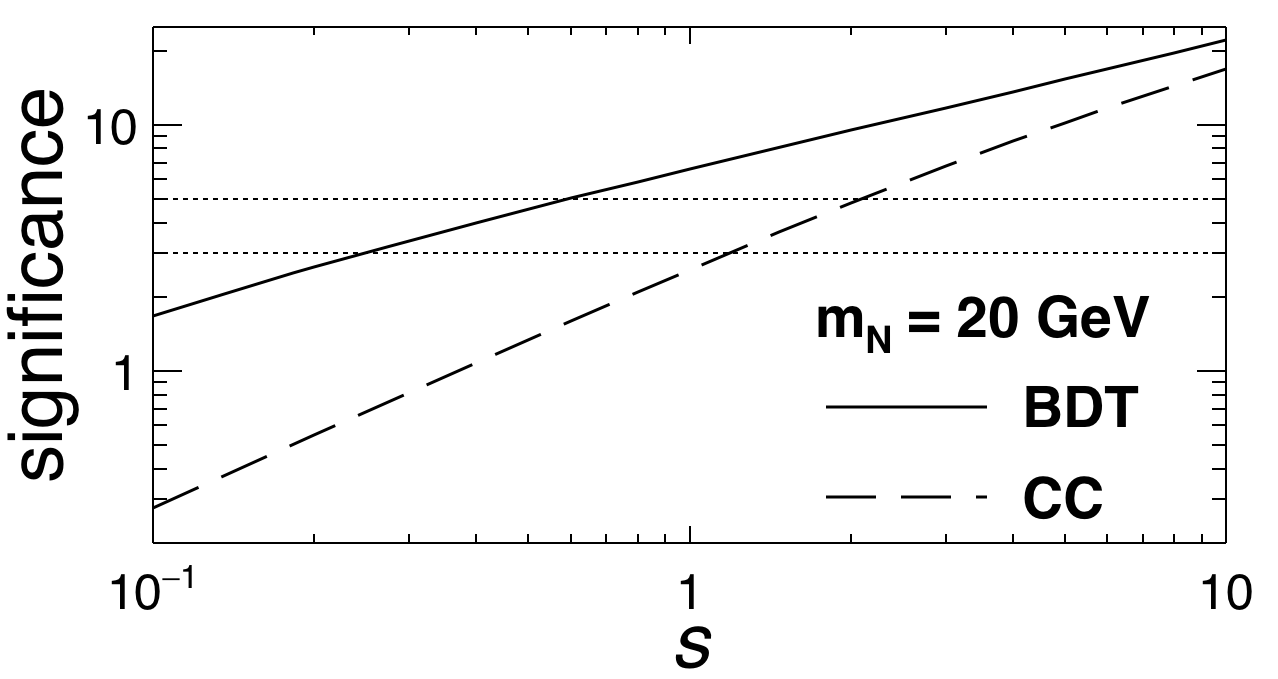}
\includegraphics[scale=0.09]{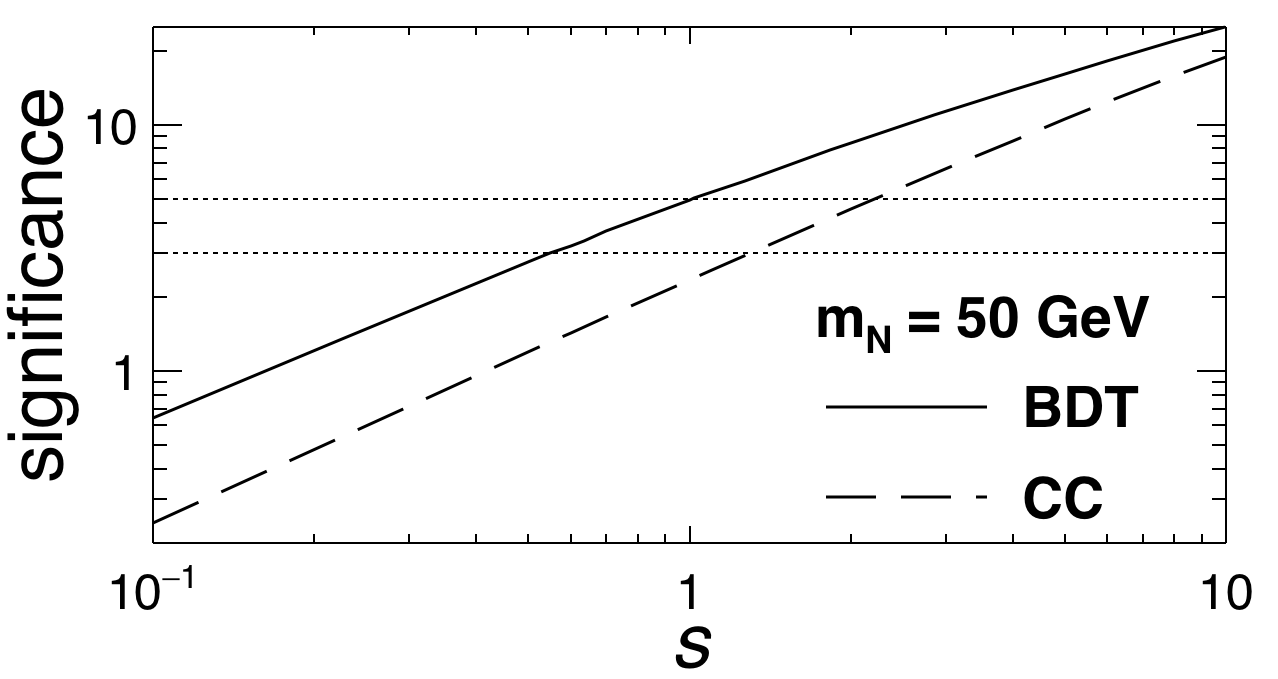}
\caption{Discovery and exclusion limits for Dirac sterile neutrinos with $m_N$ = 20 (left) and 50 (right) GeV. }
\label{fig:sgfDrc}
\end{figure}

Fig.~\ref{fig:sgfDrc}
shows the discovery and exclusion curves for a Dirac $N$, for both the BDT and CC methods.
By exploiting more useful kinematical observables and better optimization compared with the CC method, the BDT method can greatly enhance the discovery and exclusion limits. Due to the small number of signal events, the performance of the BDT method becomes close to that of CC method for small $s$ values (see Table~\ref{tab:Nsig}). Using the BDT method, one can get significances $ \geq 5.0 \sigma\, (3.0 \sigma) $   for  $s \geq 0.55\, (0.25) $ at $m_N = $ 20 GeV, or $s \geq 1.02\, (0.55)$ at $m_N = $ 50 GeV.

\begin{figure}[h]
\includegraphics[scale=0.09]{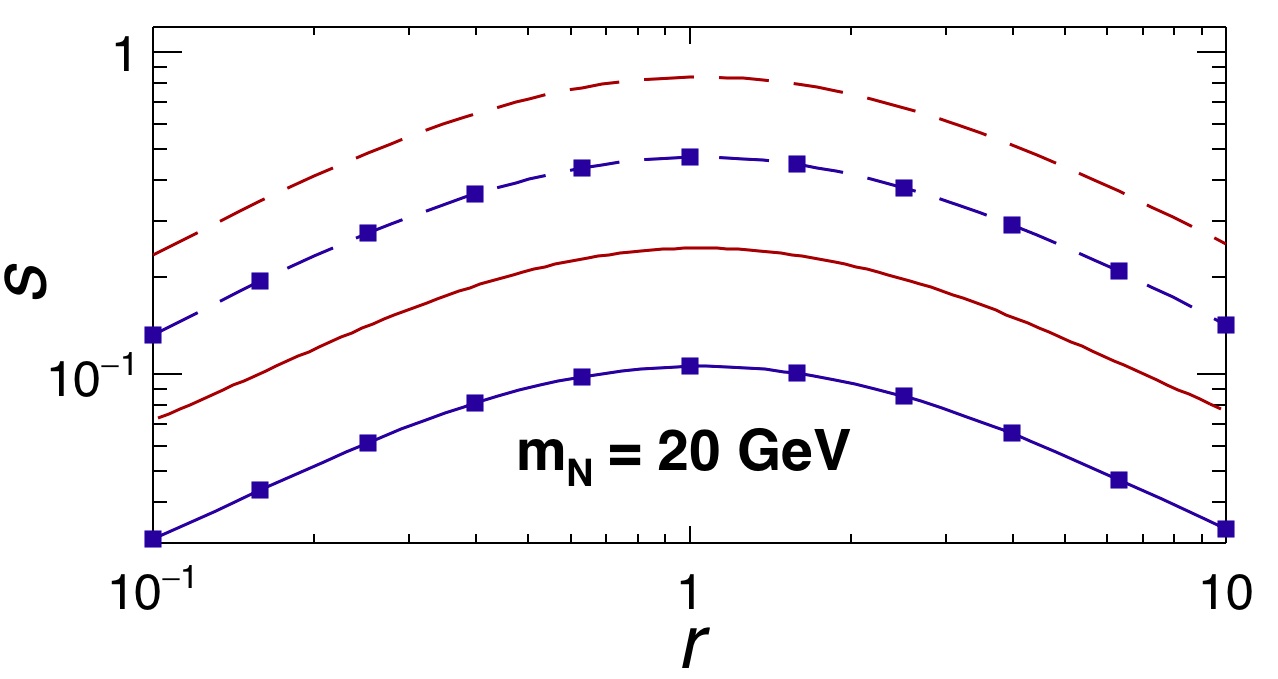}
\includegraphics[scale=0.09]{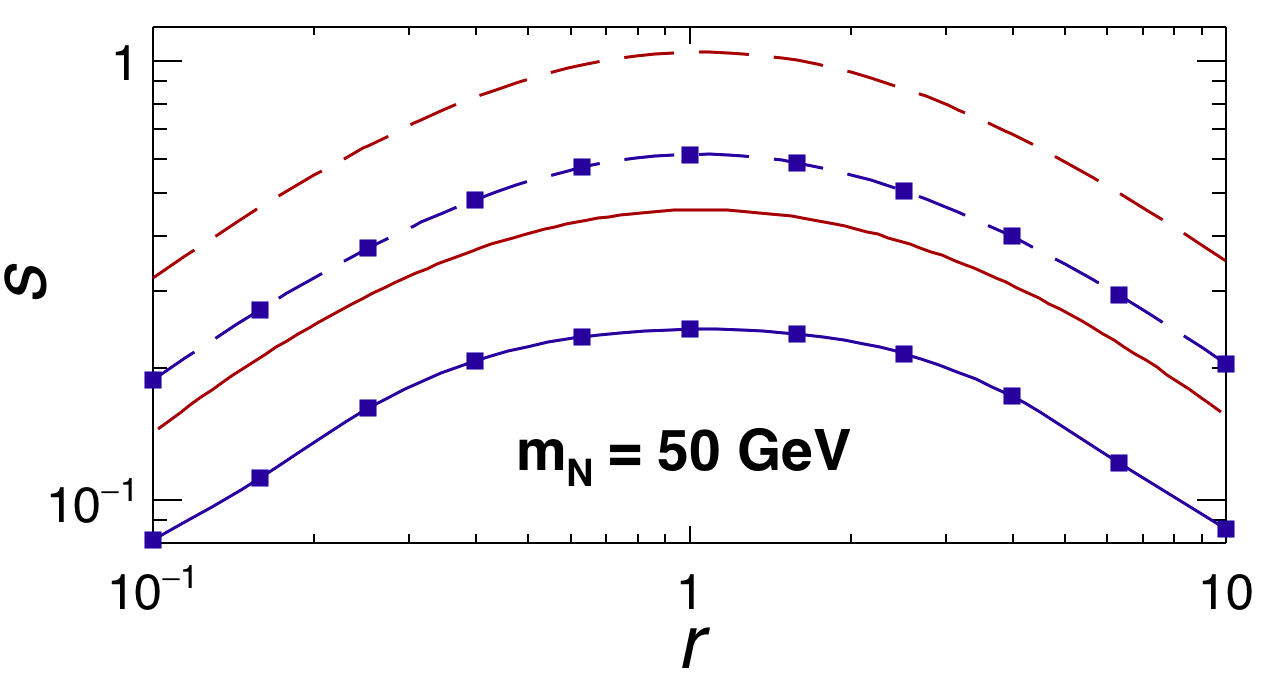}
\caption{Discovery and exclusion limits for Majorana sterile neutrinos with $m_N$ = 20 (left) and 50 (right) GeV, where the blue curves marked with squares correspond to 3-$\sigma$ limit, while the red curves correspond to 5-$\sigma$ limit; solid lines for BDT method and dashed lines for CC method.}
\label{fig:sgfMaj}
\end{figure}

Fig.~\ref{fig:sgfMaj}
shows the discovery and exclusion curves for a Majorana $N$, using both the BDT and CC methods. Here the rates depend on both $s$ and $r$ (see Table~\ref{tab:Nsig}), and so the observables at the LHC can be used to constrain both $s$ and $r$.  When $r = 1$, one can get  a significance above $ 5.0\sigma \, (3.0 \sigma)$ for $s \geq 0.24\, (0.11) $ at $m_N = $ 20 GeV, or  $s \geq 0.46\, (0.25) $ at $m_N = $ 50 GeV. For a given $s$,  the significance becomes larger when $r \neq1$, due to the larger number of signal events. Using the BDT method, when $r \approx$ 10, one can get significances $ \geq 5.0\sigma \, (3.0 \sigma)$  for $s \geq 0.08\, (0.03) $ at $m_N = $ 20 GeV, or $s \geq 0.16\, (0.09) $ at $m_N = $ 50 GeV.
\\ 

\noindent{\bf Discrimination Limit:} We now show that one can distinguish between a Dirac and Majorana $N$ in the trilepton events,  using the following distributions, which differ between the LNC and LNV processes: ($i$) the transverse mass of the system formed by the missing energy plus lepton(s) $M_T(\met, l_N)$, $M_T(\met, {l^\prime}_N)$, and $M_T(\met, {l^\prime}_N l_W )$; and ($ii$) the azimuthal angle difference $\Delta\phi$ between the missing transverse momentum and lepton(s) $\Delta\phi(\met, l_N)$,
$\Delta\phi(\met, {l^\prime}_N)$, and $\Delta\phi(\met, {l^\prime}_N l_W )$.

In order to exploit these differences, we must first reduce as much SM background as possible: after applying the basic cuts and vetoes, we perform the first BDT analysis and input the rest of the observables except those mentioned in the above paragraph to suppress the SM backgrounds. Simulated Majorana data are input as the signal sample, while the total SM background data are input as the background sample for TMVA training and testing processes.
After the first BDT cut, the total number of events, for $M_N = 20$ GeV, including all four final states ($e^\pm e^\pm\mu^\mp$ and $\mu^\pm\mu^\pm e^\mp$)
for the Dirac signals (the LNC rate only), Majorana signals (LNC + LNV rates) and SM backgrounds ($\gamma^*/Z$, $W^{\pm} Z$, and $t\bar{t}$ inclusively) are 48.5, 120.4 and 7.3, respectively.

Since $s$ is a global scale a priori unknown,
as a second step we adjust $s$ for the Dirac hypothesis to match the number of events of the Majorana hypothesis, so that our simulation does not artificially distinguish the two scenarios simply by the rates.
Just as in Ref.~\cite{Dib:2016wge}, the best matched value of $s_{\rm D}$ is found by minimizing:
\begin{eqnarray}
\chi^2_{H} = -2\, \underset{s}{\text{min}} \left \{ \text{ln} \left( \prod_i ~ \text{Poiss} \left [ N_i^{\text{expc}}, N_i^{\text{obs}}(s) \right ] \right) \right \},
\end{eqnarray}
where $i$ indicates a particular trilepton final state,  Poiss($N^{\rm expc}$, $N^{\rm obs}$) denotes the probability of observing $N^{\rm obs}$ events in Poisson statistics when the number of expected events is $N^{\rm expc}$.
Here $N^{\rm expc}$ is the expected number of events for the Majorana hypothesis (LNC + LNV + SM background), while $N^{\rm obs}$ is the observed number of events for the Dirac hypothesis (LNC + SM background).
The best matched $s_{\rm D}$ found in this way for the Dirac hypothesis gives the closest number of events to the Majorana case. For $m_N$ = 20 GeV, we find $s_{\rm D}\sim 2.44$. After matching, the Dirac and Majorana hypotheses will have 125.6 and 127.6 events, respectively.

\begin{figure}[h]
\includegraphics[scale=0.09]{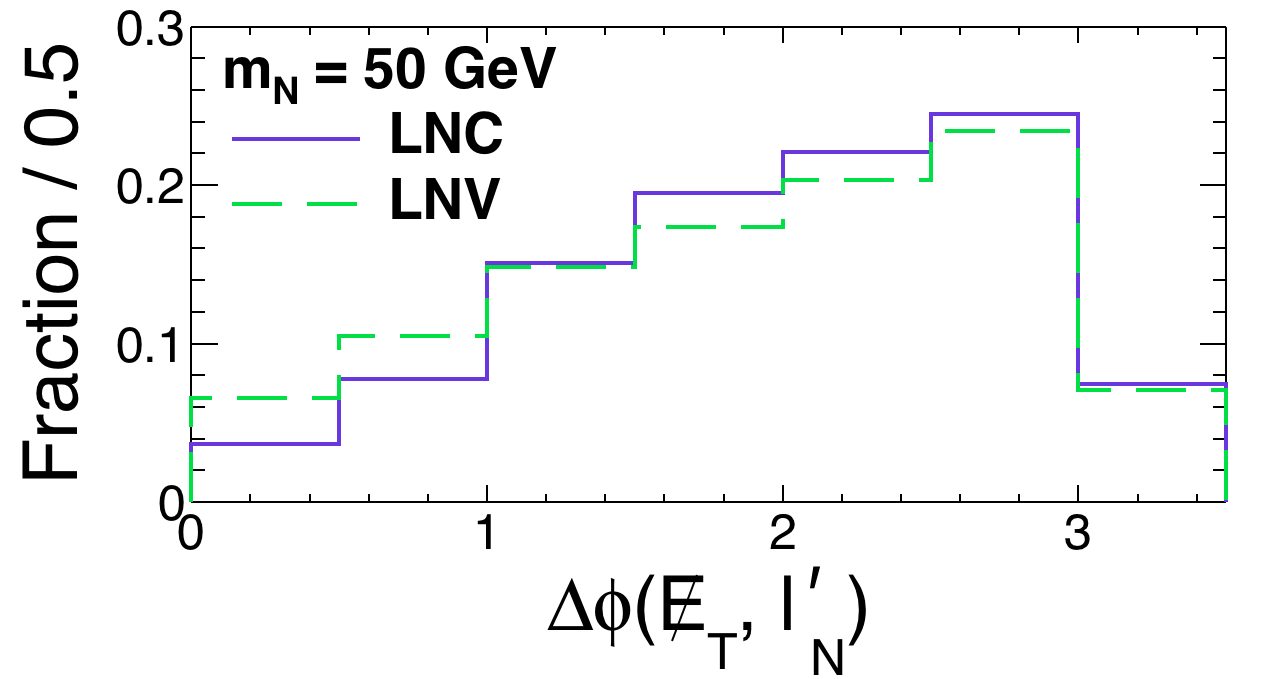}
\includegraphics[scale=0.09]{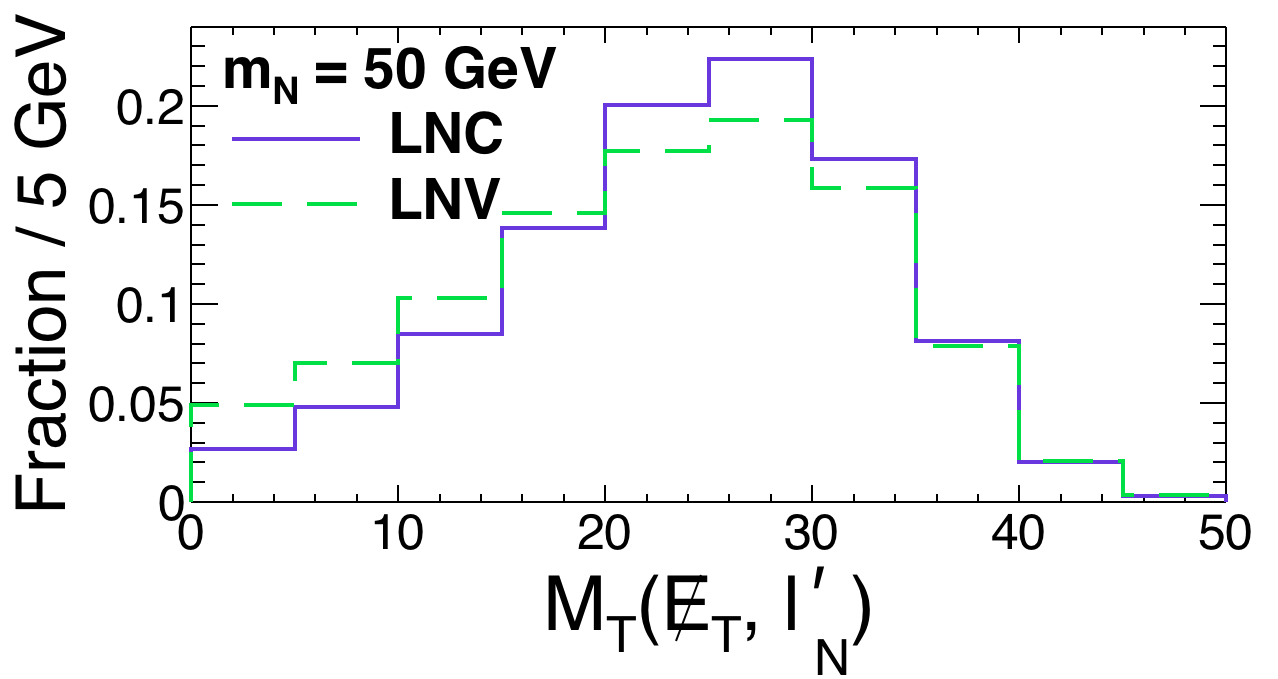}
\caption{Distributions for the benchmark point $m_N$ = 50 GeV after applying the basic cuts, b-jets veto and the first BDT cut.}
\label{fig:obsLNCvsLNV}
\end{figure}
As a third step, we perform a second BDT analysis to distinguish Majorana from Dirac hypothesis by exploiting the differences in the distributions, mentioned above. Fig.~\ref{fig:obsLNCvsLNV} shows the distributions of two of these observables after basic cuts, b-jets veto and the first BDT cut.
With an optimized second BDT cut of about 0.020, the Majorana case ends up with 46.1 events, while the Dirac hypothesis has 34.1 events. After defining
the excess in the Majorana case from the Dirac hypothesis as the ``signal" events $N_s$, and the number of events of the Dirac hypothesis as the ``background" events $N_b$, the significance for distinguishing Majorana from Dirac can be calculated as $\mathit{s} = N_s/\sqrt{N_s+N_b} = (46.1 - 34.1)/\sqrt{46.1} \approx 1.8$. This three-step method can be extended to the case where $r \neq 1 $.

\begin{figure}[b]
\includegraphics[scale=0.09]{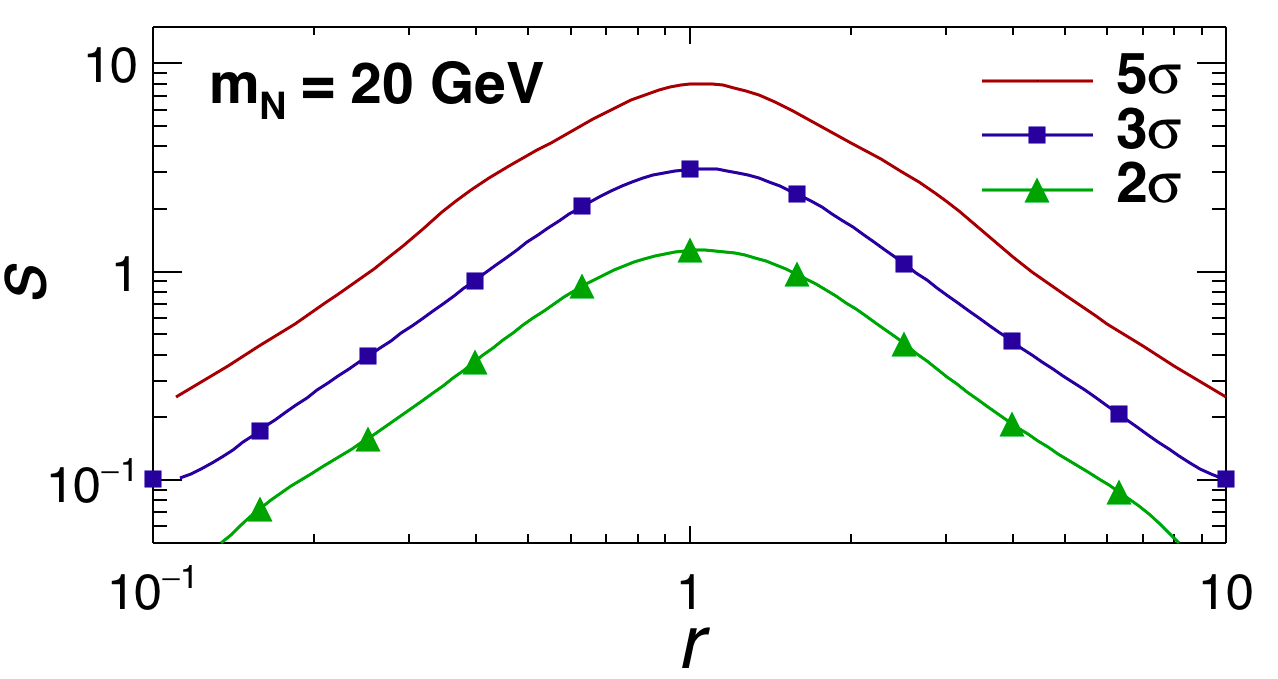}
\includegraphics[scale=0.09]{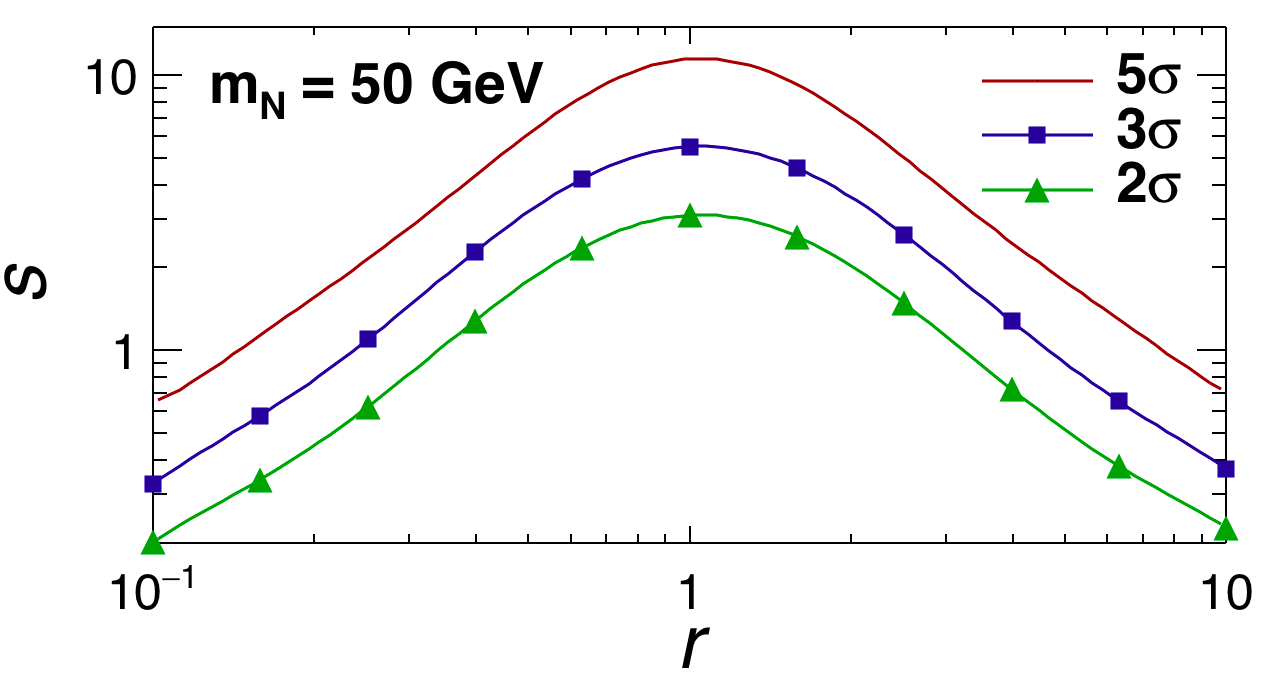}
\caption{Confidence levels of distinguishing between Dirac and Majorana neutrinos for $m_N$ = 20 (left) and 50 (right) GeV. }
\label{fig:sgfDrcVSmaj}
\end{figure}
When $r \neq 1$, the number of events for different trilepton states will be quite different between Dirac and Majorana
(see Table~\ref{tab:Nsig}), which helps in this discrimination and gives a higher significance.
Fig.~\ref{fig:sgfDrcVSmaj} shows the Confidence levels for distinguishing  Majorana from Dirac after the above three-step method. When $r \approx 1$, one can have significances $\geq 5.0\sigma (3.0 \sigma)$ for $s \geq 7.93 (3.10)$ at $m_N = 20$ GeV, or $s \geq 11.44 (5.47)$ at $m_N = 50$ GeV. As $r \approx$ 10, the same significance is reached with lower
$s \sim 0.25 (0.10)$ at $m_N = 20$ GeV, or 0.72 (0.38) at $m_N = 50$ GeV.
\\

\noindent{\bf Summary} We present a complete method to discover or set exclusion limits for heavy sterile neutrinos with $m_N < M_W$, and discriminate  their Dirac vs. Majorana nature, in trilepton final states at the 14 TeV LHC, using both Cut-and-Count (CC) and Multi-Variate Analysis (MVA) methods.
Expressing the mixings in terms of $s$ and $r$ [c.f. Eq.~(\ref{sr})],  for an integrated luminosity of $3000~\text{fb}^{-1}$,
using the MVA method, a significance of 5.0 (0.3)$\sigma$ can be achieved when $s \geq 0.55 (0.25)$ for a Dirac sterile neutrino with $m_N = $ 20 GeV, or $s \geq 1.02 (0.55)$ with $m_N = $ 50 GeV. Here we would like to recall that, according to Eq.~(\ref{eqn:mixings}), when $r=1$, the mixings are $|U_{Ne}|^2=|U_{N\mu}|^2=s\times 10^{-6}$.

For Majorana sterile neutrinos, the same significances can be reached when $r \approx$ 10, $s \geq 0.08 (0.03)$ for $m_N = 20$ GeV, or $s \geq 0.16 (0.09)$ for $m_N = $ 50 GeV. Let us recall that, when $r=10$, the mixings are $|U_{Ne}|^2=10\,|U_{N\mu}|^2=5.5\, s\times 10^{-6}$.

Moreover, Majorana vs. Dirac can be distinguished with those significances when $r \approx 1$ and $s \geq 7.9 (3.1)  $ for $m_N = 20$ GeV, or $s \geq 11 (5.8)$ for $m_N = 50$ GeV. As $r \approx$ 10, the same significances are reached  for $s\geq 0.25 (0.10)$ for $m_N = 20$ GeV, or $s\geq 0.72 (0.38)$ for $m_N = 50$ GeV.

Therefore, for an integrated luminosity of 3000 ${\rm fb}^{-1}$ at the 14 TeV LHC, both Dirac and Majorana sterile neutrinos can be found with $5 \sigma$ significance if heavy-to-light neutrino mixings $|U_{Ne}|^2 \sim |U_{N\mu}|^2\sim 10^{-6}$, while Majorana vs. Dirac discrimination can be reached if at least one of the mixings is of order $10^{-5}$.

\begin{acknowledgments}
We thank Jue Zhang for his valuable help. K.W. was supported by the International Postdoctoral Exchange Fellowship Program (No.90 Document of OCPC, 2015);  C.S.K. by the NRF grant funded by the Korean government of the MEST (No. 2016R1D1A1A02936965);
and C.D. by Chile grants Fondecyt No.~1130617, Conicyt ACT 1406 and PIA/Basal FB0821.
\end{acknowledgments}


\nocite{*}

\begin{thebibliography}{99}

\bibitem{Oscillation_experiment} Y.~Fukuda {\it et al.} [Super-Kamiokande Collaboration],
    Phys.\ Rev.\ Lett.\  {\bf 81}, 1562 (1998), 
    hep-ex/9807003; Q.~R.~Ahmad {\it et al.} [SNO Collaboration],
  \emph{ibid.} {\bf 89}, 011301 (2002), 
  nucl-ex/0204008.

\bibitem{seesaw_models}
J.W.F. Valle and J.C.~Romao,
\emph{Neutrinos in high energy and astroparticle physics},
ISBN-13: 978-3527411979
(1st Edition, Wiley-VCH, Berlin, 2015).

\bibitem{Majorana_theory} E.~Majorana,
  Nuovo Cimento  {\bf 14}, 171 (1937); 
  G.~Racah,
  \emph{ibid.} {\bf 14}, 322 (1937). 

\bibitem{Engel:2016xgb}
  J.~Engel and J.~Men\' endez, (2016),
  arXiv:1610.06548;


\bibitem{Majorana_experiments} H.~V.~Klapdor-Kleingrothaus {\it et al.},
Eur.\ Phys.\ J.\ A {\bf 12}, 147 (2001),
hep-ph/0103062; A.~M.~Bakalyarov {\it et al.} [C03-06-23.1 Collaboration],
Phys.\ Part.\ Nucl.\ Lett.\  {\bf 2}, 77 (2005) [Pisma Fiz.\ Elem.\ Chast.\
Atom.\ Yadra {\bf 2005},
21 (2005)], hep-ex/0309016;
H.~V.~Klapdor-Kleingrothaus and I.~V.~Krivosheina,
Mod.\ Phys.\ Lett.\ A {\bf 21}, 1547 (2006); M.~Agostini {\it et al.} [GERDA Collaboration],
Phys.\ Rev.\ Lett.\  {\bf 111},
122503 (2013), %
arXiv:1307.4720;
A.~Pocar [EXO-200 and nEXO Collaborations],
Nucl.\ Part.\ Phys.\ Proc.\  {\bf 265-266}, 42 (2015);
Y.~Gando [KamLAND-Zen Collaboration],
Nucl.\ Part.\ Phys.\ Proc.\  {\bf 273-275}, 1842 (2016);
A.~Gando {\it et al.} [KamLAND-Zen Collaboration],
Phys.\ Rev.\ Lett.\  {\bf 117},
082503 (2016);
{\bf 117},
109903 (2016),  %
arXiv:1605.02889.

\bibitem{Deppisch:2015qwa} 
  F.~F.~Deppisch, P.~S.~Bhupal Dev and A.~Pilaftsis,
  New J.\ Phys.\  {\bf 17}, no. 7, 075019 (2015)
  doi:10.1088/1367-2630/17/7/075019
  [arXiv:1502.06541 [hep-ph]].

\bibitem{Antusch:2016ejd} 
  S.~Antusch, E.~Cazzato and O.~Fischer,
  Int.\ J.\ Mod.\ Phys.\ A {\bf 32}, no. 14, 1750078 (2017)
  doi:10.1142/S0217751X17500786
  [arXiv:1612.02728 [hep-ph]].

\bibitem{dileptonLHC}
G.~Aad {\it et al.} [ATLAS Collaboration],
JHEP {\bf 1507}, 162 (2015), 
arXiv:1506.06020;
V.~Khachatryan {\it et al.} [CMS Collaboration],
Phys.\ Lett.\ B {\bf 748}, 144 (2015), 
arXiv:1501.05566.


\bibitem{Izaguirre:2015pga}
  E.~Izaguirre and B.~Shuve,
  Phys.\ Rev.\ D {\bf 91},
  093010 (2015),
  arXiv:1504.02470.


\bibitem{p2mu}
  G.~Cvetic, C.O.~Dib and C.S.~Kim,
  JHEP {\bf 1206}, 149 (2012),
  arXiv:1203.0573.

\bibitem{Anamiati:2016uxp} 
  G.~Anamiati, M.~Hirsch and E.~Nardi,
  JHEP {\bf 1610}, 010 (2016),
  [arXiv:1607.05641 [hep-ph]].


\bibitem{Dib:2015oka}
  C.O.~Dib and C.~S.~Kim,
  Phys.\ Rev.\ D {\bf 92},
  093009 (2015),
  arXiv:1509.05981.


\bibitem{Dib:2016wge}
  C.O.~Dib, C.S.~Kim, K.~Wang and J.~Zhang,
  Phys.\ Rev.\ D {\bf 94},
  013005 (2016),
  arXiv:1605.01123.

\bibitem{Helo:2013esa}
  J.~C.~Helo, M.~Hirsch and S.~Kovalenko,
  Phys.\ Rev.\ D {\bf 89}, 073005 (2014);
 {\bf 93},
  099902(E) (2016),
  arXiv:1312.2900.

\bibitem{Dib:2014iga}
  C.O.~Dib and C.~S.~Kim,
  Phys.\ Rev.\ D {\bf 89},
  077301 (2014),
  arXiv:1403.1985.

\bibitem{Antusch:2016vyf} 
  S.~Antusch, E.~Cazzato and O.~Fischer,
  JHEP {\bf 1612}, 007 (2016)
  doi:10.1007/JHEP12(2016)007
  [arXiv:1604.02420 [hep-ph]].

\bibitem{TMVA2007}
  P.~Speckmayer, A.~Hocker, J.~Stelzer and H.~Voss,
  J.\ Phys.\ Conf.\ Ser.\  {\bf 219}, 032057 (2010);
A.~Hocker {\it et al.},
  PoS ACAT, 040 (2007),
  physics/0703039.


\end{thebibliography}
\end{document}